# Quantum States of Higher-order Whispering gallery modes in a Silicon Micro-disk Resonator

RAKESH RANJAN KUMAR, YI WANG, YAOJING ZHANG AND HON KI TSANG*

*Department of Electronic Engineering, The Chinese University of Hong Kong, Shatin NT, Hong Kong SAR*
*email: hktsang@ee.cuhk.edu.hk*

**Abstract:** The quantum states of light in an integrated photonics platform provide an important resource for quantum information processing and takes advantage of the scalability and practicality of silicon photonics. Integrated resonators have been well explored in classical and quantum optics. However, to encode multiple information through integrated quantum optics requires broader utilization of the available degrees of freedom on a chip. Here, we studied the quantum interference between photon pairs of the same higher order whispering gallery modes populated by spontaneous four-wave mixing in an integrated silicon micro-disk resonator. The quantum interference between the photon pairs of the first two quasi-TE$_0$ and quasi-TE$_1$ radial modes was measured to be $V_{net} \sim 98 + 0.8$ % and $V_{net} \sim 94 + 2.6$ %, respectively. The results are promising for achieving higher-dimensional quantum states using the higher-order radial modes of a micro-disk resonator coupled with an integrated waveguide.

## 1. Introduction

Integrated quantum photonics have been extensively explored for quantum information processing and quantum technologies [1-3]. Silicon quantum photonics offers a small footprint, high nonlinearity, scalability, CMOS compatibility, low-cost production, low power operation and mature fabrication technique promising the future practical quantum information processing on a chip [1, 4-6]. Therefore, exploring the higher dimensionality of the existing quantum photonic sources on the silicon platform would allow us to encode more information per photon [7]. One approach was demonstrated to encode information using the higher degree of freedom in free-space and fiber-based quantum sources to produce qudits and higher dimensional entanglement states [8-9]. In integrated quantum optics, recent experiments using quantum interference of the transverse-modes in a multi-mode waveguide in the silicon nitride platform demonstrated its potential for use in quantum information processing [10, 11]. Transverse-mode entanglement between higher-order modes was also reported using the silicon-on-insulator (SOI) platform [12]. A multi-mode waveguide can support many co-propagating modes, which can be used as parallel channels to encode information through single photons in a more compact way for future quantum information processing due to their orthogonality relation between the propagating modes [11].

Chip scale entangled photon-pair sources have already been demonstrated in silicon micro-ring and micro-disk resonators which are the backbone for efficient quantum information processing [6, 13-17]. However, the quantum interference between the photon pairs of the same transverse modes for the higher degree of freedom has never been exploited previously in resonators which produce bright correlated photon pairs [16].

In this paper, we produce non-classical light from the higher-order whispering gallery modes (WGM) in a micro-disk resonator with on-chip integrated waveguide for coupling. Although multi-mode micro-disk resonators are well-known in classical optics [18], the fully integrated micro-disk resonators have not been explored previously for integrated quantum photonics. The quantum interference visibility from the photon pairs produced from the first transverse mode was measured to be > 98 %. The different radial orders of the resonator give rise to different resonance peaks in the spectral response of the micro-disk resonator and the generation of entangled photons in the same order radial modes through quantum frequency combs. Higher-dimensional frequency-bin entangled quantum states in micro-ring resonators were demonstrated by optical frequency modulation to mix the quantum frequency combs lines [19, 20]. However, a large cavity resonator is needed to ensure that the frequency spacing could be addressed by the modulation at 50 GHz [20]. The potential advantage of using different radial orders WGM are that they can provide resonances which are closely spaced in frequency, in small volume (high Purcell factor) resonators and these closely spaced resonances can be used for generating higher-order qudit states without needing the use of the large cavities resonators used in ref. [20]. The different radial order whispering gallery modes can be very closely separated from each other, with some modes separated by only a few GHz, and this reduces the modulation frequency speed needed for mixing the frequency-bin quantum entangled states.

## 2. Device Fabrication and Characterization

The micro-disk resonator schematic is shown in Fig. 1 (a). The micro-disk resonator was fabricated using electron-beam lithography to define the device layout on the SOI wafer with 250 nm top silicon layer and a 3 μm buried-oxide ($SiO_2$). Fig. 1(a) shows the fully etched silicon waveguide which was produced by dry etching using $C_4F_8$ and $SF_6$ gases. The quasi TE mode profiles were simulated using an eigen-mode solver (Lumerical MODE solutions) as illustrated in Fig. 1(b-d) for the first three quasi TE modes of the micro-disk resonator. Single-mode optical fibers were used for in/out coupling via waveguide grating couplers which have losses of about 8-dB per fiber-waveguide interface and a center wavelength at 1565 nm.

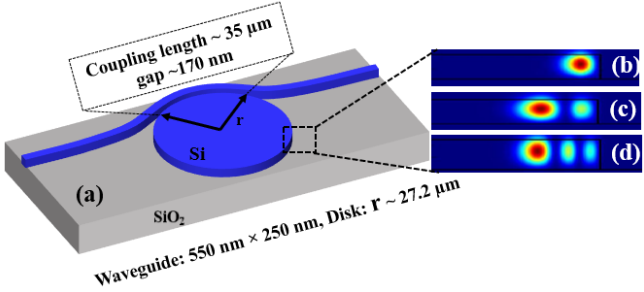

Fig. 1. (a) Micro-disk resonator schematic and dimension (b-d) Depict the first three transverse electric (TE) modes.

The device transmission spectrum was measured by a continuous wave (CW) tunable laser, supporting fundamental and higher order modes which can be seen in Fig. 2(a). The spectra contain different higher order resonances labelled as $TE_0$ and $TE_1$ modes which have equal mode spacing and thus satisfy the energy conservation requirement for spontaneous four-wave mixing (SFWM). The wavelengths difference among higher-order modes is ~ 0.1 nm approximately, as shown below in Fig. 2 (b). The loaded quality factors (Q) were shown for the $TE_0$ and $TE_1$ modes of the pump, signal and idler wavelengths in Fig. 2(b).

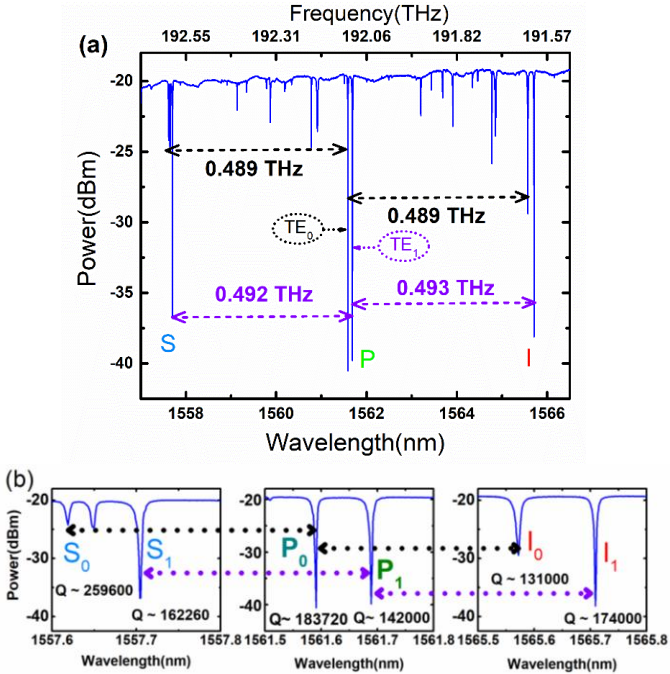

Fig. 2(a) Transmission spectrum of Si Micro-disk resonator with $TE_0$ and $TE_1$ modes were shown with equal frequency spacing. (b) Zoom in plot of the signal, pump and idler resonance modes from (a) for the $TE_0$ and $TE_1$ modes.

A small frequency detuning of the radial modes could be adjusted in the device layout design by changing the pulley waveguide coupling length and the varying a gap between pulley waveguide and the resonator. The micro-disk resonator was mounted on a temperature controller stage with thermo-electric cooler feedback with a sensor with accuracy maintained at ± 0.02 °C, to make it stable with pump wavelength on a resonance wavelength. The correlated pair of photons were routed through a 50:50 fiber beam splitter in order to non-deterministically split the signal and idler photons. The total pump power rejection is estimated to be over 120 dB. The total optical loss on signal and idler wavelengths was ~ 16 to 18 dB, including the 8-dB coupler loss of the output waveguide-fiber interface.

## 3. Results

### 3.1 Photon-pair generation by Whispering gallery modes

In SFWM, two pump photons are annihilated, creating signal and idler photons which are quantum correlated and must satisfy the energy conservation ($2\omega_p = \omega_s + \omega_i$) as shown in Fig. 2(a) by equal frequency detuning from the pump wavelength [21]. The resonator can greatly enhance the internal generation rate depending on the Purcell enhancement (Q/V) [22]. The internal generation rate scales as $R \propto Q^3 P^2 r^{-2}$, where Q is the quality factor of the resonances, r is the radius of the resonators, and P is the coupled power in the waveguide [23].

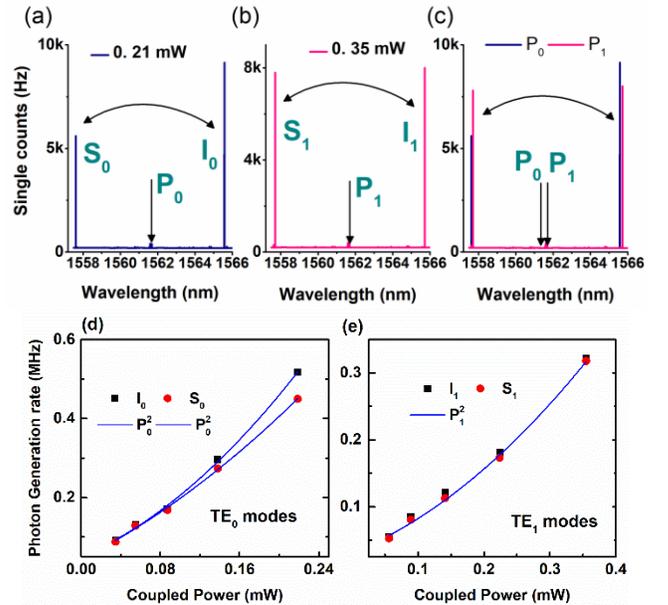

Fig. 3 (a) SWFM was performed on $TE_0$ and (b) $TE_1$ modes independently (c) When both pumps are used, SFWM process occur independently and measured single photon counts were shown in the vertical axis for each case. (d-e) The photon generation rate for the $TE_0$ and TE$_1$ modes vs coupled power inside the cavity scales as P$_0^2$ or P$_1^2$ and fitted with quadratically by blue-line. The photon generation rates were plotted after accounting the total coupling loss in the output and the detection efficiency.

Cavity-enhanced SFWM was performed by pumping at the resonance wavelengths 1561.59 nm (P$_0$) or 1561.65 nm (P$_1$) and signal and idler photons were collected at 1557.65 nm or 1557.70 nm and 1565.57 or 1565.71 nm, respectively as shown in Fig. 3(a-b). The raw single photons count rates for the TE$_0$ modes were measured to be ~ 5.6 kHz and 9.1 kHz at the signal (S$_0$) and idler (I$_0$) wavelength, respectively, for the coupled power of 0.21 mW as shown in Fig. 3(a). Similarly, for TE$_1$ modes, the measured raw single-photon count rates were 7.9 kHz and 8.0 kHz at the signal (S$_1$) and idler (I$_1$) wavelength respectively, at 0.35 mW coupled power is shown in Fig. 3(b). Finally, we use two pump P$_0$ and P$_1$ to excite both TE$_0$ and TE$_1$ modes simultaneously as shown in Fig. 3 (c), the counts observed in both detectors were the sum of individuals counts in the first two-plot in Fig. 3 (a-b). Hence, the two pump P$_0$ and P$_1$ contribute the SFWM process independently inside the cavity, and there is no correlation between the different higher radial orders because they are spatially orthogonal. Therefore, higher-order radial modes inside the cavity can be used to encode the parallel quantum information in an efficient/compact way compare to multimode waveguide [11,12,16].

The spectral brightness of the photon pairs produced at TE$_0$ and TE$_1$ modes are 1.92×10$^7$ pair/s/mW$^2$/GHz and 7.30×10$^6$ pair/s/mW$^2$/GHz, respectively and it is slightly lower than the reported in taper-fiber coupled micro-disk resonator [16]. The spectral brightness of the source is slightly lower because of the lower quality factor of the radial modes compare to ref. [16], but comparable to another on-chip integrated waveguide coupled micro-disk resonator [24].

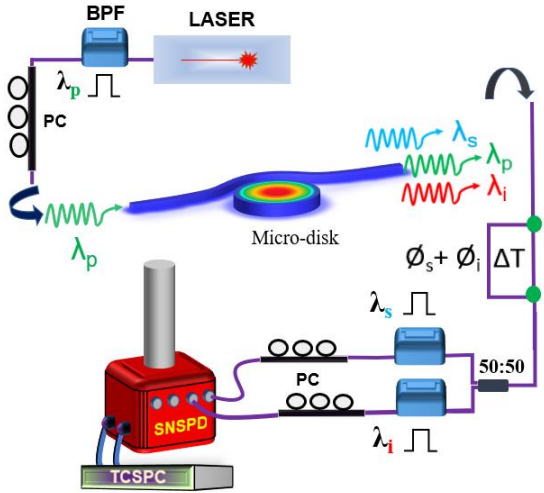

Fig. 4. Experimental set-up: BPF: Band pass filter, PC: polarization controller, SNSPD: Superconducting nanowire single photon detector, TCSPC: Time correlated single photon counting. In Actual experiment two laser were combined with the 50:50 fiber beam splitter to obtain the experimental results as shown in the Fig. 3 (c).

One of the major reasons why the micro-disk resonator reported in this paper has lower spectral brightness because of the much larger radius decreases the Purcell enhancement (Q/V) inside the cavity compared to the previously reported work [6, 16, 23]. The larger size of the disk suffers from higher dispersion inside the cavity and higher-order modes can be seen in Fig. 2(a) which are not tightly confined like a micro-ring resonator leads to expand the mode volume larger inside the cavity for higher-order radial modes results in a lower Purcell enhancement.

### 3.2 Coincidence counts

After observing the bright pair of photons using WGM, we characterized the quantum correlation between the photon pairs by the coincidence-to-accidental ratio (CAR) which is commonly used as a figure of merit of the source [6, 25]. The CAR value was obtained from the time-bin histogram measurement (Fig. 5(a-b)) without an unbalanced Mach-Zehnder interferometer (UMZI) reported in the experimental set-up in Fig. 4. The measured CAR value for TE$_0$ and TE$_1$ modes is 439 ± 30 and 108 ± 42 at the photon generation rate of 9.1 kHz and 8.0 kHz in Fig. 4(a-b), respectively. These CAR values are much higher than those obtained in a multimode waveguide [12]. The CAR value was calculated from the number of coincidences at the peak over the number of coincidences away from the peak in Fig. 5(a, b) and error bars represent the accidental count variations at different delays. However, this quantity mainly depends on the coupled power in the waveguide and can be further improved if losses were minimized at the output of the device to the detector [6, 16]. The exact instant of a photon pair emission from the cavity with a CW pump laser is uncertain. Therefore, quantum-correlated photons can be characterized by measuring the temporal correlations of photon detection events between the signal and idler channels [26].

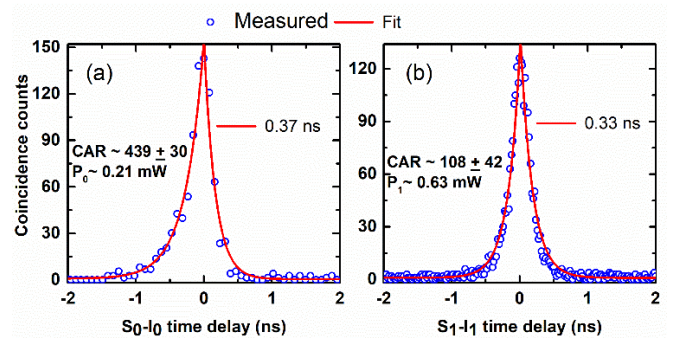

Fig. 5. Photon-pair verifications by time-bin histogram width (a) coincidence counts histogram of S$_0$-I$_0$ modes at 0.21 mW coupled power in the waveguide. The integration time to obtain the histogram was 360 seconds (b) Coincidence counts histogram of S$_1$-I$_1$ modes at 0.63 mW coupled power in the waveguide. The integration time to obtain the histogram was 480 seconds.

To characterize the orthogonal photon pairs emitted from the micro-disk resonator at the TE$_0$ and TE$_1$ modes, we used exponential fits to the above histogram plots on either-side to determine the bandwidth of the resonator [27]. The experimental data in Fig. 5(a-b) fitted exponentially to each side to determine the decay times of the signal and idler photons. The width of the above histogram is related to the coherence time of the emitted photons corresponding to the bandwidth of the resonator and combined time jitter of the superconducting single-photon detectors (SNSPD, detection efficiency~ 70 %, dark count < 300 Hz) and the time tagging electronics [16]. The fitted decay time constants of the signal ($S_0$) and idler ($I_0$) photons were 210 ps and 110 ps for the $TE_0$ modes which agree well with theoretically calculated from the Q-factor ($S_0 \sim$ 215 ps and $I_0 \sim$ 109 ps) of the corresponding resonance modes. Similarly, For the $TE_1$ modes, the fitted decay time constant of the signal ($S_1$) and idler ($I_1$) photons were 136 ps and 139 ps, respectively. The theoretically predicted value from the Q-factor of the $S_1$ and $I_1$ modes was 134 ps and 144 ps, respectively. Hence, we have demonstrated that the SFWM process occurs independently inside the cavity at $TE_0$ and $TE_1$ modes which can be used for parallel quantum information processing [11, 12].

### 3.3 Time-Energy Entanglement test on $TE_0$ and $TE_1$ modes

To assess the time-energy entanglement on photon pairs emitted from the WGM, we have used a folded fiber Franson interferometer configuration which consists of a single UMZI due to the ease of implementation and robustness against system instability [17]. The fiber interferometer used in the experiment was mounted on copper mounted Peltier stage sealed in a metal box which is completely isolated from any environmental fluctuations and has precise control of the phase by a temperature controller at accuracy maintained at $\pm$ 0.01°C. In order to observe the high visibility time-energy entanglement, the UMZI must satisfy the condition ($\tau_c << \Delta T << \tau_p$) first proposed by Franson [28]. Where $\tau_c \sim$ 215 ps, is bi-photon coherence time (extracted from the linewidth of one of the resonance modes), $\Delta T \sim$ 1.8 ns, the unbalanced arm of the interferometer and $\tau_p \sim$ 100 μs, is coherence time of the CW laser in our measurement set-up, clearly satisfying the proposed requirement. Notice that the time-energy entanglement verification was performed using a single UMZI as shown in Fig. 4 which is differs from Franson's proposal but serves the purpose of characterizing the time-energy entanglement on photon pair generated from the micro-disk resonator [17].

The correlated photons have passed through the single UMZI and a 50:50 fiber beam splitter non-deterministically separate the signal and idler photons as illustrated in Fig. 4. The correlated-pair of photons within the coherence time of the pump laser are passed through UMZI following either the same path (short-short or long-long) or different path (short-long or long-short), resulting in three peaks in the coincidence histogram separated by the unbalanced arm delay ($\Delta T \sim$ 1.8 ns) as shown in the Fig. 6(c-e). The side peaks in the histogram are distinguishable due to their difference in arrival times while the middle peaks are clearly indistinguishable as both photons either travels from long-long or short-short arm, hiding their path information. Hence, these photons can be quantum correlated and expected to oscillate as a function of cosine $C(\varphi) \propto \cos(\varphi + \delta)$ with the interferometer phase. where $\phi = \phi_s + \phi_i$ is the sum of the phase of the signal and idler photons and $\delta$ is a constant phase term [17,29].

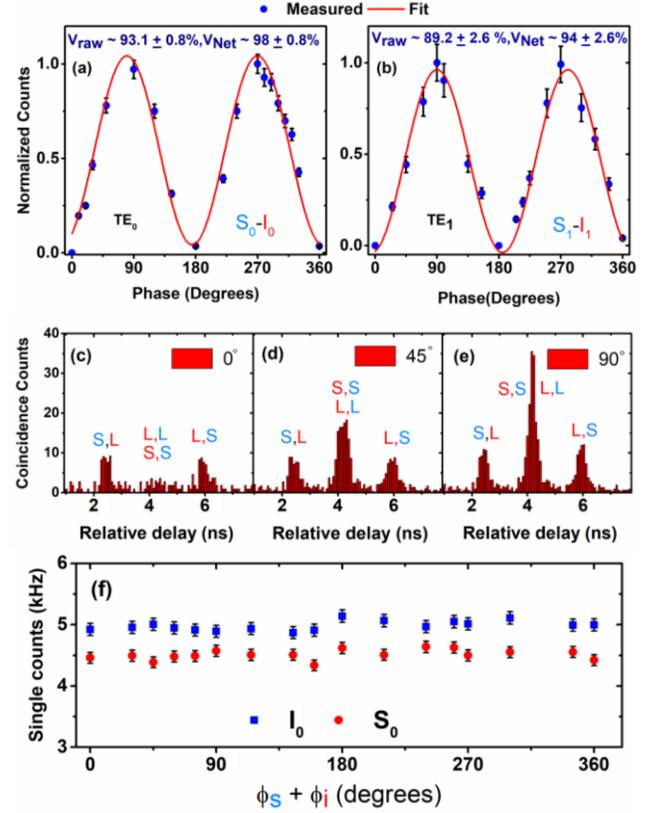

Fig. 6 Two-photon interference fringes from the time-energy entanglement (a) Normalized coincidence counts from $S_0$-$I_0$ resonance modes vs phase of the UMZI at coupled power 0.21 mW. (b) Normalized coincidence counts from $S_1$-$I_1$ resonance modes vs UMZI at coupled power 0.63 mW. (c)-(e) Raw coincidence counts with the phase of UMZI at 0, 45 and 90 degrees in Fig. 6 (a). The error-bars are represented by the one standard deviation. (f) The absence of single counts interference with the interferometer phase clearly indicates that the interferometer unbalanced arm is much greater than the emission time of the bi-photon coherence time.

We have observed the raw two-photon quantum interference visibilities for the $TE_0$ and $TE_1$ is 93.1 $\pm$ 0.8 % at the coupled power of 0.21 mW and 89.2 $\pm$ 2.6 % at the coupled power of 0.63 mW, respectively. After subtracting the background counts, we obtained the interference visibilities of 98 $\pm$ 0.8 % and 94 $\pm$ 2.6 % for the $TE_0$ and $TE_1$ modes shown in Fig. 6(a-b) which is greater than $\sim$ 71 %, thus confirming the entanglement through violation of Bell's inequality [30]. The measured visibility is defined as $V_{Meas} = (N_{max} - N_{min})/(N_{max} + N_{min})$, Where $N_{max}$ and $N_{min}$ represent the maxima and minima of the peak in Fig. 6(c-e), obtained from the best fit of the experimental data and error on the visibility is represented by the one standard deviation. In Fig 6(c-e), the middle peak transforms from a maximum to close to zero allows us to measure the high visibility entangled photon pairs while side peaks are remained to be unaffected by UMZI.

### 4. Discussion

The micro-disk resonator reported in this paper is CMOS compatible and uses an integrated bus waveguide (instead of the tapered fiber [16, 17] or prism [27] for evanescent coupling). To the best of my knowledge, this is the first time such an integrated system is used as a source in the field of quantum optics. The micro-disk can support higher-order transverse mode analogous to multi-mode waveguides, and the photon pair generated exhibits excellent spectral brightness that can align well with other quantum photonic devices [10-17, 22-24]. The higher-order transverse modes can be used for high-capacity quantum information processing as well as parallel quantum information processing through each mode to encode the quantum bit information in a compact way. In this work, we measured the quantum interference of the photon pairs generated from the first two transverse mode in a resonator, but it is possible to be extended other higher-order modes. Scaling to more modes for more quantum states is essential for quantum computation and can be achieved by increasing the width of integrated waveguide [31, 32]. We have not demonstrated the intermodal interaction ($TE_0TE_1$) between the transverse modes in this work. In the future separation of the different bus waveguide modes with the use of asymmetrical directional coupler can separate the $TE_0$ and $TE_1$ modes [12] or using four-port microdisk resonator which allows us to resolve $TE_0$ and $TE_1$ in a separate port. Notice that we have used pulley coupled waveguide which allows extraction of the photons from the cavity in an efficient way for each mode. However, the gap between the resonator and waveguide must be engineering well in order to increase the efficiency of the photon pairs from each transverse mode. In our case, we have chosen the gap in such a way that the $TE_1$ mode to be near critical coupling which automatically increases the extraction efficiency of the $TE_0$ modes as well.

The possibility of achieving 100 % visibility from the correlated photon pairs depends on the ability to precisely control the phase of UMZI and measurement stability. Any imperfections in the UMZI may add the artifact in the measured coincidence counts which reduces the oscillation of the observed peak-value. The sensitivity to environment disturbances of the fiber based UMZI used in our experiments is one contribution to the lower visibility of ∼ 98 %.

## 5. Conclusion

In conclusion, we have reported the high visibility entanglement of $TE_0$ and $TE_1$ modes from a silicon micro-disk resonator for the first time coupled with on-chip integrated waveguide operating at the room temperature which can be used for generating higher-dimensional quantum states. The net visibility reaches for $TE_0$ and $TE_1$ modes are 98 and 94 %, respectively, which is comparable with the best reported earlier on silicon nanophotonic chip-scale sources [6, 17]. The spectral brightness of our source is the same order of magnitude as reported in fiber-coupled micro-disk resonator [16]. The main advantage of using the device proposed here is its fully monolithic implementation, unlike the use of the tapered fiber coupler and prism coupler of previous work [16, 17, 27]. We also showed that the different radial order modes of WGM can produce closely spaced frequency resonances. These can facilitate the generation of higher order qudit states when modulated to mix the different radial modes [19, 20]. Our scheme proposed here can be applied for even higher-order TE modes and not just limited to the first two TE modes as reported which can exponentially increase the number of qubit states required for the quantum computation and information processing [7-12].

**Disclosures.** The authors declare no conflicts of interest.

**Funding Information.** Hong Kong Research Grants Council (RGC), General Research Fund (GRF) (14207117).

**Acknowledgment**. We would like to thank Xiankai Sun, Jingwen Ma, Zeije Yu and Ning Zhang and Chester Shu for sharing their equipment.

**References**

1. J. Wang, Fabio Sciarrino, A. Laing, and Mark G. Thompson, "Integrated Photonic quantum technologies," Nat. Photonics **7**, 6404, (2019).
2. B. J. Metcalf et al., "Quantum teleportation on a photonic chip," Nat. Photonics **8**, 770 (2014).
3. J. B. Spring et. al, "Boson sampling on a photonic chip," Science **339**, 798 (2013).
4. N. Matsuda, H. L. Jeannic, H. Fukuda, T. Tsuchizawa, W. John Munro, Kaoru Shimizu, K. Yamada, Y. Tokura and H. Takeuse, "Monolithically integrated polarization entangled photon pair source on a silicon chip," Sci. rep. **2**, 817 (2012).
5. J. W. Silverstone, R. Santagati, D. Bonneau, M. J. Strain, M. Sorel, J. L. O'Brien and M. G. Thompson, "Qubit entanglement between ring-resonator photon-pair sources on a silicon chip. Nat. Comm. **6**, 7948 (2015).
6. C. Ma, X. Wang, V. Anant, A. D. Beyer, M. D. Shaw, Z. and S. Mookherjea, "Silicon photonic entangled photon-pair and heralded single photon generation with CAR > 12000 and $g^2(0) < 0.006$," Opt. Express. **25**, 32995 (2017).
7. B. P. Lanyon, Marco Barbieri, M. P. Almeida, T. Jennewein, T. C. Ralph, K. J. Resch, G. J. Pryde, J. L. O'Brien, A. Gilchrist and Andrew G. White, "Simplifying quantum logic using higher dimensional Hilbert spaces," Nat. Phys. **5**, 757 (2009).
8. A. Mair, A. Vaziri, G. Weihs and A. Zeilinger, "Entanglement of the orbital angular momentum states of photons," Nature **412**, 313 (2001).
9. Ebrahim Karimi, D. Giovannini, E. Bolduc, N. Bent, Filippo M. Miatto, Miles J. Padgett, and R. W. Boyd, "Exploring the quantum nature of the radial degree of freedom of a photon via Hong-Ou-Mandel interference," Phys. Rev. A. **89**, 013829 (2014).
10. L. T. Feng, Ming Zhang, Zhi-yuan Zhou, M. Li, X. Xiong, L. Yu, Bao-shen Shi, Guo-ping Guo, D. X. Dai, Xi-feng Ren, and G-can Guo, "On-chip coherent conversion of photonic quantum entanglement between different degrees of freedom," Nat. comm. **7**, 11985 (2016).
11. A. Mohanty, M. Zhang, A. Dutt, S. Ramelow, P. Nussenzveig, and M. Lipson, "Quantum interference between transverse spatial waveguide modes," Nat. Comm. **8**, 14010 (2017).
12. Lan-Tian Feng, Ming Zhang, Xiao Xiong, Y. Chen, H. Wu, M. LI, Guo-ping Guo, Guan-can Guo, Dao-xin Dai, and Xi-Feng Ren, "On-chip Transverse-mode entangled photon pair source," npj Quantum Inf. **5,** 2 (2019).
13. S. Azzini, D. Grassani, Michael J. Strain, Marc Sorel, L.G. Helt, J.E. Sipe, M. Liscidini, M. Galli, D. Bajoni, "Ultra-low power generation of twin photons in a compact silicon ring resonator," Opt. Express **20**, 23100 (2012).
14. D. Grassani, S. Azzini, M. Liscidini, M. Galli, M. J. Strain, M. Sorel, J. E. Sipe and D. Bajoni, "Micrometer-scale integrated silicon source of time-energy entangled photons," Optica **2**, 88 (2015).

15. R. Wakabayashi, M. Fujiwara, K. Yoshino, Y. Nambu, M. Sasaki, and T. Aoki, "Time-bin entangled photon pair generation from si micro-ring resonator," Opt. Express **23**, 1103 (2015).
16. W. C. Jiang, X. Lu, J. Zhang, O. Painter and Q. Lin, "Silicon-chip source of bright photon pairs," Opt. Express **23**, 20884 (2015).
17. S. Rogers, D. Mulkey, X. Lu, W.C. Jiang and Q. Lin, "High visibility Time-energy entangled photons from a silicon nanophotonic chip," ACS photonics **3**, 1754 (2016).
18. Dmitry V. Strekalov, C. Marquardt, A.B. Matsko, H.G.L. Schewefel, and G. Leuchs, "Nonlinear and quantum optics with whispering gallery resonators," J. Opt. **18**, 123002, (2016).
19. M. Kues, C. Reimer, P. Roztocki, L.R. Cortes, S. Sciara, B. Wetzel, Y. Zhang, A. Cino, S. T. Chu, B. E. Little, D. J. Moss, L. Caspani, J. Azana, and R. Morandotti, "On-chip generation of high-dimensional entangled quantum states and their coherent control," Nature **546**, 622 (2017).
20. P. Imany, Jose A. Jaramillo-Villegas, O. D. Odele, Kyunghun Han, Daniel E. Leaird, J. M. Lukens, P. Lougovski, M. Qi and A. M. Weiner, "50-GHz-spaced comb of high-dimensional frequency-bin entangled photons from an on-chip silicon nitride microresonator," Opt. Express **26**, 1825 (2018).
21. Robert W. Boyd, Nonlinear Opt., 3rd ed., New York (2008).
22. S. Clemmen, K.P. Huy, W. Bogaerts, R. Baets, P. Emplit, and S. Massar, "continuous wave photon pair generation in silicon-on-insulator waveguides and ring resonators," Opt. Express **17**, 16558 (2009).
23. S. Azzini, D. Grassani, M. Galli, L.C. Andreani, M.Sorel, M. J. Strain, L.G. Helt, J. E. Sipe, M. Liscidini and D. Bajoni, Super spontaneous four-wave mixing in single-channel side-coupled integrated spaced sequence of resonator structures," Opt. Lett. **37**, 4431 (2012).
24. R. R. Kumar, X. Wu, H. K. Tsang, "Photon pair generation and filtering using monolithically integrated silicon micro-disk and coupled resonator optical waveguide," CLEO, STh1H.2 (2019).
25. O. Alibart, J. Fulconis, G. K. L. Wong, S. G. Murdoch, W. J. Wadsworth, and J. G. Rarity, "Photon pair generation using four-wave mixing in a microstructured fibre: theory versus experiment," New J. Phys. **8**, 67 (2006).
26. Roy J. Glauber, "The quantum theory of optical coherence" Phys. Rev. **130**, 2529 (1963).
27. M. Fortsch, Josef U. Furst, C. Wittman, D. Strekalov, A. Aiello, M. V. Chekhova, C. Silberhorn, Gerd Leuchs, and Christoph Marquardt, "A versatile source of single photons for quantum information processing," Nat. Comm. **4**, 1818 (2013).
28. J.D. Franson, "Bell inequality for position and time" Phys. Rev. Lett. **62,** 2205 (1989).
29. R. R. Kumar, M. Raevskaia, V. Pogoretskii, Y. Jiao, and H. K. Tsang, "Entangled photon-pair generation from an InP membrane micro-ring resonator," Appl. Phys. Lett. **114**, 021104 (2019).
30. P.G. Kwiat, A.M. Steinberg, and R.Y. Chiao, "High-Visibility interference in a Bell-inequality experiment for energy and time," Phys. Rev. A **47**, R2472 (1993).
31. M. Zukowski, A. Zellinger, Michael A. Horne, "Realizable higher-dimensional two-particle entanglements via multiport beam splitters," Phys. Rev. A **55**, 2564 (1997).
32. D. Dai, C. Li, S. Wang, H. Wu, Y. Shi, Z. Wu, S. Gao, T. Dai, H. Yu and H. K. Tsang, "10-channel mode de-multiplexer with dual polarizations," Laser Photonics Rev. **12**, 1700109 (2018).